\pdfoutput=1
\documentclass{elsarticle}

\usepackage{lineno}
\usepackage{hyperref}
\usepackage[utf8]{inputenc}
\usepackage{alphabeta}
\usepackage{amssymb}
\usepackage{amsmath}
\usepackage[english]{babel}
\usepackage{graphicx} 
\usepackage{wrapfig} 
\graphicspath{ {images/} }

\usepackage{array}
\newcolumntype{P}[1]{>{\centering\arraybackslash}p{#1}}

\usepackage{color}

\begin{document}

\begin{frontmatter}

\title{Performance of the First 150 mm Diameter Cryogenic Silicon Ionization Detectors with Contact-Free Electrodes}

\author[UMN]{N. Mast}
\author[UMN]{A. Kennedy}
\author[UMN]{H. Chagani}
\author[UMN]{D. Codoluto}
\author[UMN]{M. Fritts}
\author[TAMU]{R. Harris}
\author[TAMU]{A. Jastram}
\author[TAMU]{R. Mahapatra}
\author[UMN]{V. Mandic}
\author[TAMU]{N. Mirabolfathi}
\author[TAMU]{M. Platt}
\author[UMN]{D. Strandberg}

\address[UMN]{School of Physics \& Astronomy, University of Minnesota, Minneapolis}
\address[TAMU]{Department of Physics and Astronomy, Texas A\&M University}

\begin{abstract}

Cryogenic semiconductor detectors operated at temperatures below 100 mK are commonly used in particle physics experiments searching for dark matter. The largest such germanium and silicon detectors, with diameters of 100 mm and thickness of 33 mm, are planned for use by the Super Cryogenic Dark Matter Search (SuperCDMS) experiment at SNOLAB, Canada. Still larger individual detectors are being investigated to scale up the sensitive mass of future experiments. We present here the first results of testing two prototype 150 mm diameter silicon ionization detectors. The detectors are 25 mm and 33 mm thick with masses 1.7 and 2.2 times larger than those currently planned for SuperCDMS.  These devices were operated with contact-free bias electrodes to minimize leakage currents which currently limit operation at high bias voltages. The results show promise for the use of such technologies in solid state cryogenic detectors.
\end{abstract}


\end{frontmatter}


\section{Motivation}

Although the effects of dark matter have been observed for over eight decades \cite{Zwicky2009}, the nature of dark matter remains unexplained. Comprising roughly 85\% of all matter \cite{Ade:2015xua}, this nonbaryonic matter has several potential candidates, with Weakly Interacting Massive Particles (WIMPs) garnering much focus by direct dark matter detection experiments. Results from these experiments as well as by the CMS \cite{Khachatryan2015} and ATLAS \cite{Aad2015} experiments at the LHC have constrained the simplest supersymmetric WIMP models, which favor 10--100 GeV/c\textsuperscript{2} WIMP masses, shifting interest to more recent models that suggest a WIMP mass below 10 GeV/c\textsuperscript{2}, such as asymmetric dark matter \cite{Cushman:2013zza}. In the SuperCDMS experiment, sensitive solid state particle detectors are used to search for the small energy deposited by WIMP-nuclear recoils. To increase sensitivity to these new WIMP parameter spaces, larger payloads of more sensitive detectors must be deployed.

One method to increase sensitivity by lowering detector energy threshold relies on the Neganov-Trofimov-Luke effect, in which work done by the applied electric field on the electron-hole pairs produced in the detector is converted into phonons along the drift paths of the charges \cite{Luke:1988,Neganov:1985khw}.  SuperCDMS has developed detectors biased at 50--100 V \cite{Agnese:2016cpb} that take advantage of Neganov-Trofimov-Luke phonon production to lower the energy threshold of the experiment, and therefore improve its sensitivity to lower mass WIMPs. However, leakage currents through the detector can reduce its sensitivity when operated in this mode, as observed at the 70 V operating bias of the CDMSlite detector \cite{Agnese:2015nto}. It has been demonstrated that high voltage biasing via vacuum-separated (contact-free) electrodes can dramatically reduce this leakage for small (56 cm\textsuperscript{3}) Ge devices \cite{Mirabolfathi:2015pha}.

The work presented here demonstrates the feasibility of such contact-free bias schemes with larger (450--580 cm\textsuperscript{3}) Si crystals.
Although these tests do not include phonon readout, they are the first demonstration of such bias schemes used with the largest cryogenic Si particle detectors yet operated at temperatures below 100 mK.

\section{Detector Designs and Experimental Setup}

\begin{figure}
	\centering
	\includegraphics[width=\textwidth]{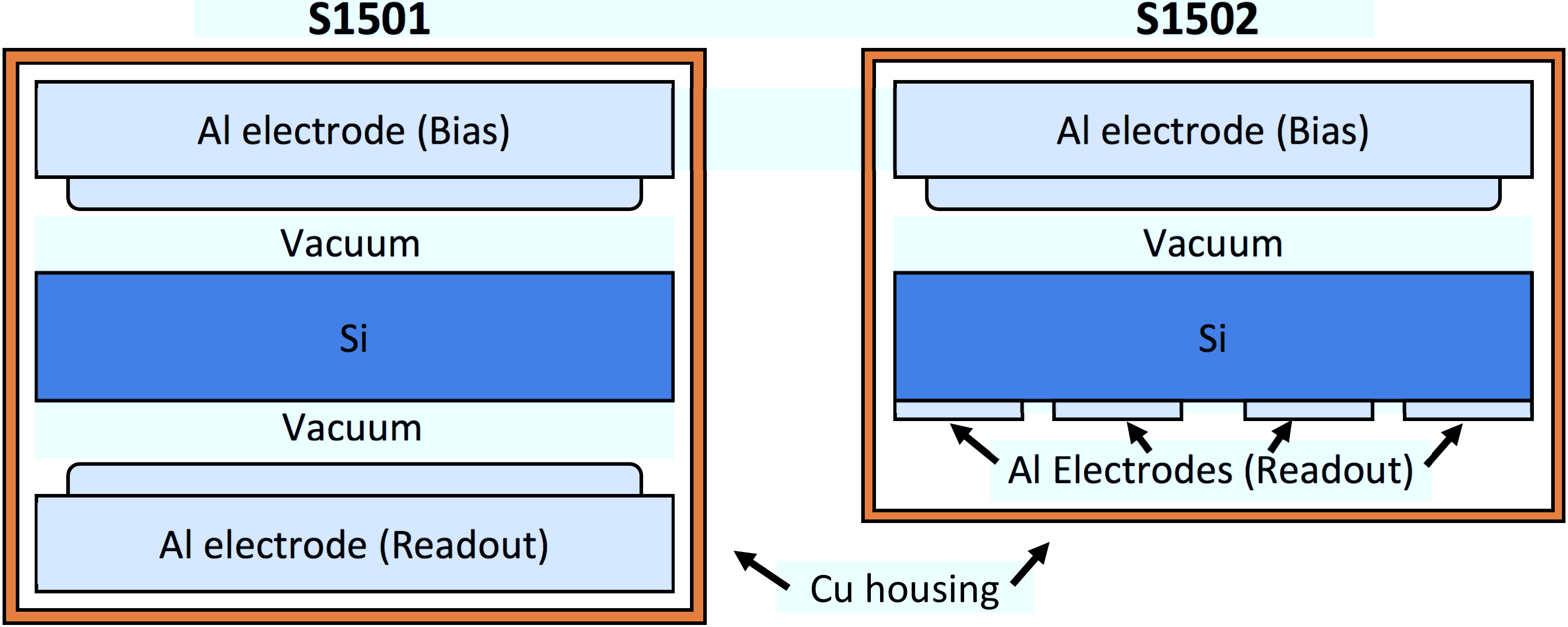}
	\caption{\textit{Left}: S1501 detector design with two aluminum electrodes (gray) surrounding a 150 mm Si crystal (blue) with a vacuum gap between faces. All pieces are secured within a copper housing with Cirlex clamps (not pictured). \textit{Right}: S1502 detector design with a single vacuum gap for biasing and a set of deposited readout electrodes on the crystal surface.} 
	\label{fig:detector_diagrams}
\end{figure}

To test the performance parameters of large diameter silicon (Si) crystals as well as the use of prototype contact-free detector designs, two ionization detectors were fabricated and tested using n-type, [100] orientation, high purity, high resistivity (\textgreater15 k\Omega~cm) Si crystals purchased from TopSil \cite{Topsil}.

One crystal, which was 150 mm diameter and 33 mm thick, was used to construct a simple single-channel contact-free detector by mounting the bare crystal between two planar aluminum electrodes. This detector is denoted ``S1501''. As shown in the left side of Fig. \ref{fig:detector_diagrams}, the ionization signal was measured on one electrode (``Readout'') while the other was used to provide a voltage bias (``Bias''). The readout electrode could also be biased up to $\pm$12 V, but most often the readout electrode was grounded and the bias electrode provided a 0--100 V bias.

\begin{figure}
    \centering
	\includegraphics[width=0.7\textwidth]{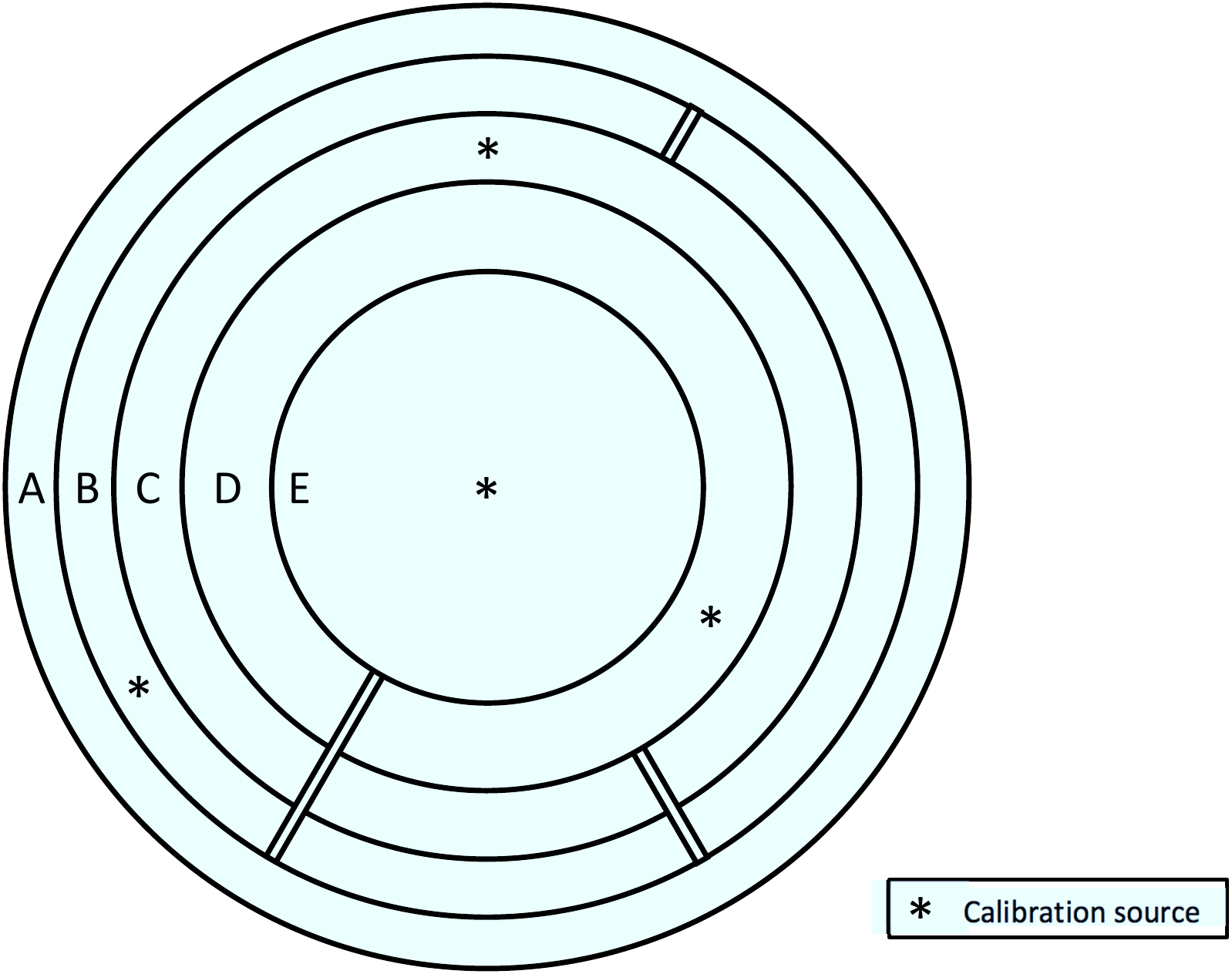}
    \caption{S1502 ionization electrode channel map. Locations of the \textsuperscript{241}Am calibration sources are marked. The segments of channels B and C are connected with wirebonds so that each forms a continuous annular electrode. Sections of the inner electrodes extend towards the detector radius to allow wirebond connections to readout electronics. The S1501 detector had a single, monolithic electrode with a single calibration source at the center.}
    \label{fig:S1502_mask}
\end{figure}

A second 150 mm diameter 25 mm thick crystal, denoted ``S1502'', had five concentric electrodes of equal areas on one crystal surface as shown in Fig. \ref{fig:S1502_mask}. The electrodes were fabricated on one crystal face by depositing a layer of amorphous Si followed by a layer of aluminum while the opposite crystal face was left bare. The five channel geometries were then defined by wet-etching. Again, a planar contact-free electrode was mounted near the bare face to provide a bias voltage and the induced ionization signals were read out from the five deposited electrode channels.

\begin{figure}
    \centering
    \includegraphics[width=0.7\textwidth]{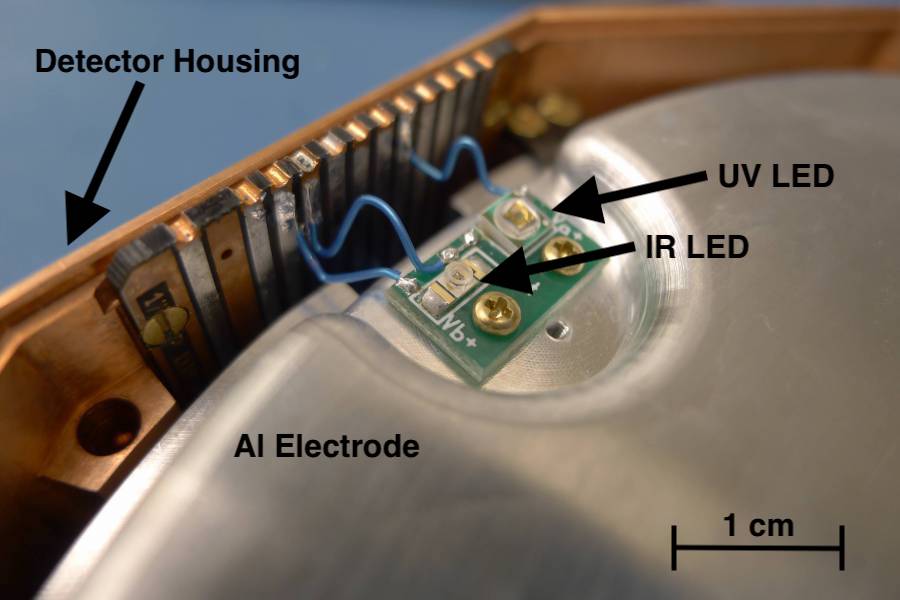}
    \caption{Photo of LED board mounted on electrode.}
    \label{fig:LED_photo}
\end{figure}

The cylindrical aluminum electrodes, shown in Fig. \ref{fig:detector_diagrams}, were comprised of a 150 mm outer diameter top lip, used to secure the electrode to its housing, and a solid 142 mm diameter cylinder which extended below the bottom edge of the lip. Each electrode was mounted with its flat face parallel to the crystal surface, with a gap of $\lesssim$1 mm. As shown in Fig. \ref{fig:LED_photo}, two shallow recesses along the radial periphery of the electrode face provided space for small circuit boards. Infrared (IR) at 940 nm and ultraviolet (UV) at 310 nm Light Emitting Diodes (LEDs) mounted on each board were used to periodically reset the detectors as discussed further in \autoref{sec:Neutralization}. The detectors were mounted and enclosed in housings machined from high purity copper. Mounting spacers were used to ensure that the contact-free electrodes and crystals were secured with the required small gaps. The voltage across the crystal, $V_{cr}$, was not the same as that applied to the detector as a whole, $V_{tot}$ due to the voltage drop across the vacuum gap(s). They can be related by

\begin{equation}
\label{eq:Vcr}
V_{cr} = \frac{h}{h+\kappa d}V_{tot}
\end{equation}
where $h$ is the detector thickness, $d$ is the total vacuum gap width, and  $\kappa=11.47$ is the relative permittivity of Si near 0 K \cite{BETHIN1974}. This ignores any fringing fields or effects of the grounded detector housing.

The vacuum gaps of S1501 were measured to be $0.29 \pm 0.03$ mm and $0.40 \pm 0.07$ mm for the readout and bias sides respectively. These gaps result in $80\% \pm 2\%$ of the total bias field being applied across the crystal itself and a total detector capacitance of $\sim$70 pF, which is comparable to typical SuperCDMS detector electrode capacitances of $\sim$100 pF \cite{Phipps:2016}. S1502 had a single gap of $0.9 \pm 0.1$ mm, giving $70\% \pm 3\%$ of applied bias across the crystal. With the different crystal thickness, vacuum gap and electrode geometry, the capacitance of each of the five readout electrodes to the bias electrode was reduced to $\sim$20 pF. Direct measurement of the vacuum gaps was difficult and is the main source of uncertainty in the magnitude of the applied electric fields and in the expected signal magnitude.

The contact-free detector studies were performed at the cryogenic detector testing facility at the University of Minnesota in an Oxford Instruments Kelvinox 100 \cite{Kelvinox} dilution refrigerator, with a base temperature ranging from 75--95 mK. The detector housings were designed to fasten to a conventional CDMS-Soudan ``tower'' \cite{Akerib:2008zz}, thus minimizing the amount of additional cold hardware to be fabricated for these tests. Signals from the readout electrodes passed from the detector to the tower and first amplifier stage via a set of coaxial wires. For S1501, the single channel detector, simple flexible coaxial cables were used. However, parasitic capacitances to ground were found to induce undesirable oscillations in the amplifier chain which distorted the pulse shapes. The design of the subsequent detector, S1502, included rigid, low capacitance vacuum coaxial connectors which eliminated this distortion and improved the data.

The standard SuperCDMS cold electronics \cite{Yvon:1996} were used to amplify the ionization signal. The LEDs and readout amplifiers were controlled and digitized with a prototype SuperCDMS Detector Control and Readout Card (DCRC) \cite{Hansen:2010} connected to a computer running MIDAS-based data acquisition software \cite{MIDAS}. The bias electrode was connected to an external power supply via a custom designed adapter board.
\section{Measured Data}

Interactions with incident particles create electron hole pairs in the crystal.  A constant bias field drifts these charges across the crystal bulk, inducing a signal on the readout electrode \cite{HE2001}, the magnitude of which can be predicted using the Shockley-Ramo Theorem \cite{Shockley,Ramo} including polarization effects of the crystal and gap dielectrics \cite{Hamel}. To first order, the expected signal at the readout electrode, $Q_{sig}$ is related to the total amount of charge liberated in the detector, $Q$ by 

\begin{equation}
Q_{sig} = \frac{h}{h+\kappa d}Q
\end{equation}
i.e. the signal is reduced by the same factor as the applied bias voltage in Eq. \ref{eq:Vcr}.

\begin{figure}

    \centering
    \includegraphics[width=\textwidth]{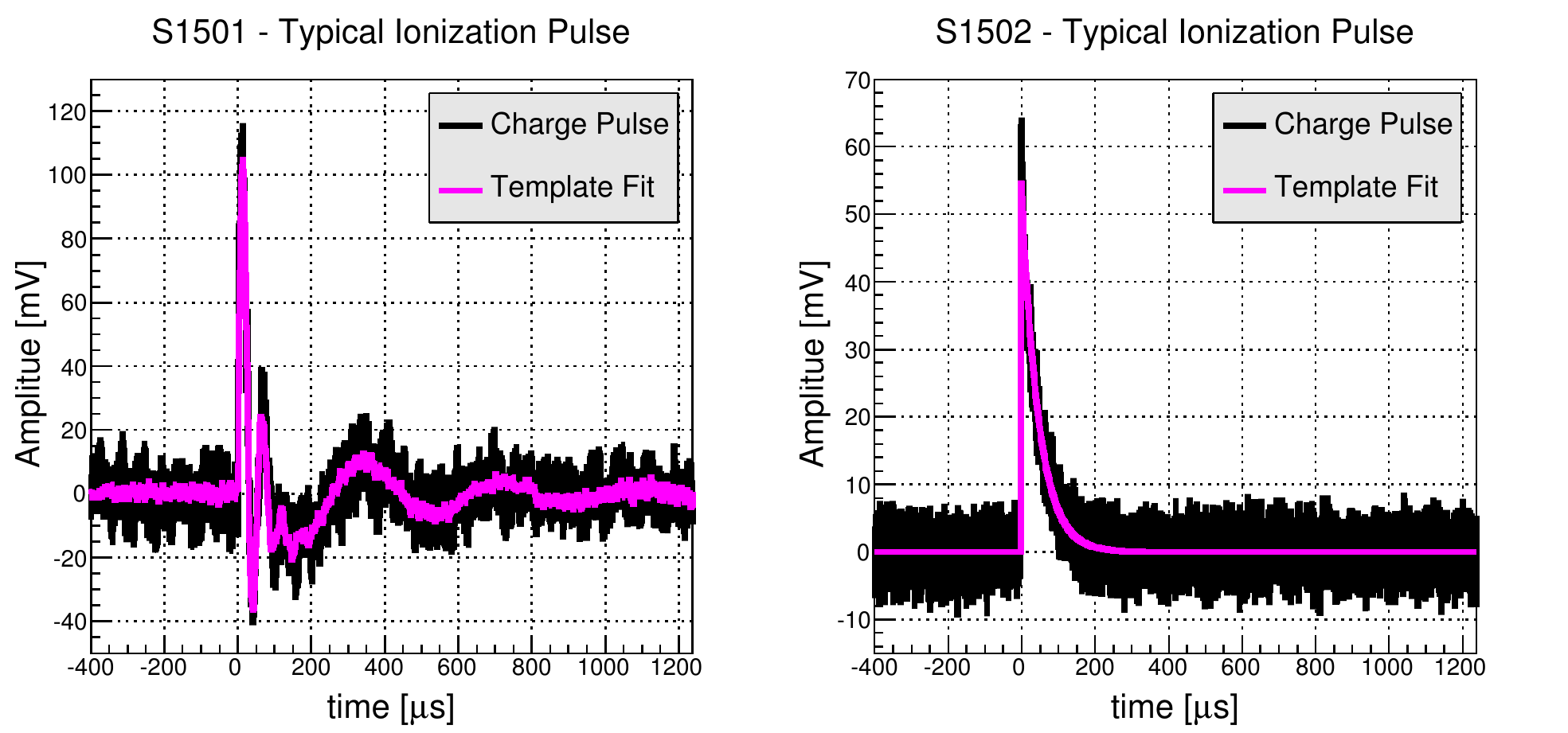}
    \caption{Typical ionization pulse signals shown in black, with fitted pulse templates in magenta. \textit{Left}: S1501 detector pulse with distortion is still well-fit by average template. \textit{Right}: S1502 channel E pulse shows effect of corrected readout circuit with simple exponential fall time.}
    \label{fig:example_pulses}
\end{figure}

The measured ionization signal is an exponential pulse with amplitude proportional to the amount of charge induced on the readout electrode.
The fall time of the pulse is set by the first stage amplifier feedback branch.  As discussed above, the coaxial cabling used with S1501 introduced a high frequency pole which caused ringing in the pulse tail (see Fig. \ref{fig:example_pulses}). Unfortunately, this prevented an absolute calibration of the S1501 readout circuit gains from first principles. However, the relative event energy information was still contained in the maximum amplitude of the pulse and the pulse shape was observed to scale linearly with signal amplitude. This allowed each pulse to be fit with an average template constructed from characteristic events, thereby estimating the relative amplitude of each event. With improved cold electronics, S1502 produced exponential pulses without ringing. This allowed an absolute calibration of the charge signals based solely on the readout circuit design for all 5 channels of S1502. In both devices, the distributions of fit $\chi^2$, time delay, and energy were used to remove poorly reconstructed events from the analysis.

\section{Experimental Results}
The 60 keV gamma line from \textsuperscript{241}Am sources was used for energy calibration. In S1501, the source was mounted at the center of the readout electrode with an 0.008" diameter hole in a 0.375" thick lead disc collimating the source along the crystal axis. In S1502, sources were mounted on the bias electrode over the center of each channel except the outermost (channel A) as shown in Fig. \ref{fig:S1502_mask}.  Each source was collimated with 0.375" thick lead discs with hole diameters of 0.018" for the center channel (E) and 0.006" for the other three (B--D). The center channel source had a different geometry and required a larger collimator hole to achieve a similar event rate to the others. The calibration events occurred more often on the side of the detector nearest the source, but because the mean free path of 60 keV gammas in Si is $\sim$3 cm \cite{NISTIR5632}, the calibration event rate only falls by about half over the depth of the crystal.

Because the sources were on opposite electrodes in these two devices, the correspondence between bias field sign and dominant charge carrier is opposite. In S1501 (S1502), the calibration sources were on the readout (bias) electrode, so positive bias field configurations preferentially involved electrons (holes) traversing most of the crystal thickness and thus generating the bulk of the induced signal. However, as explained above, this is a weak effect due to the relative uniformity of calibration event depths.

\subsection{Charge Collection Efficiency}
\begin{figure}
    \centering
    \includegraphics[width=0.8\textwidth]{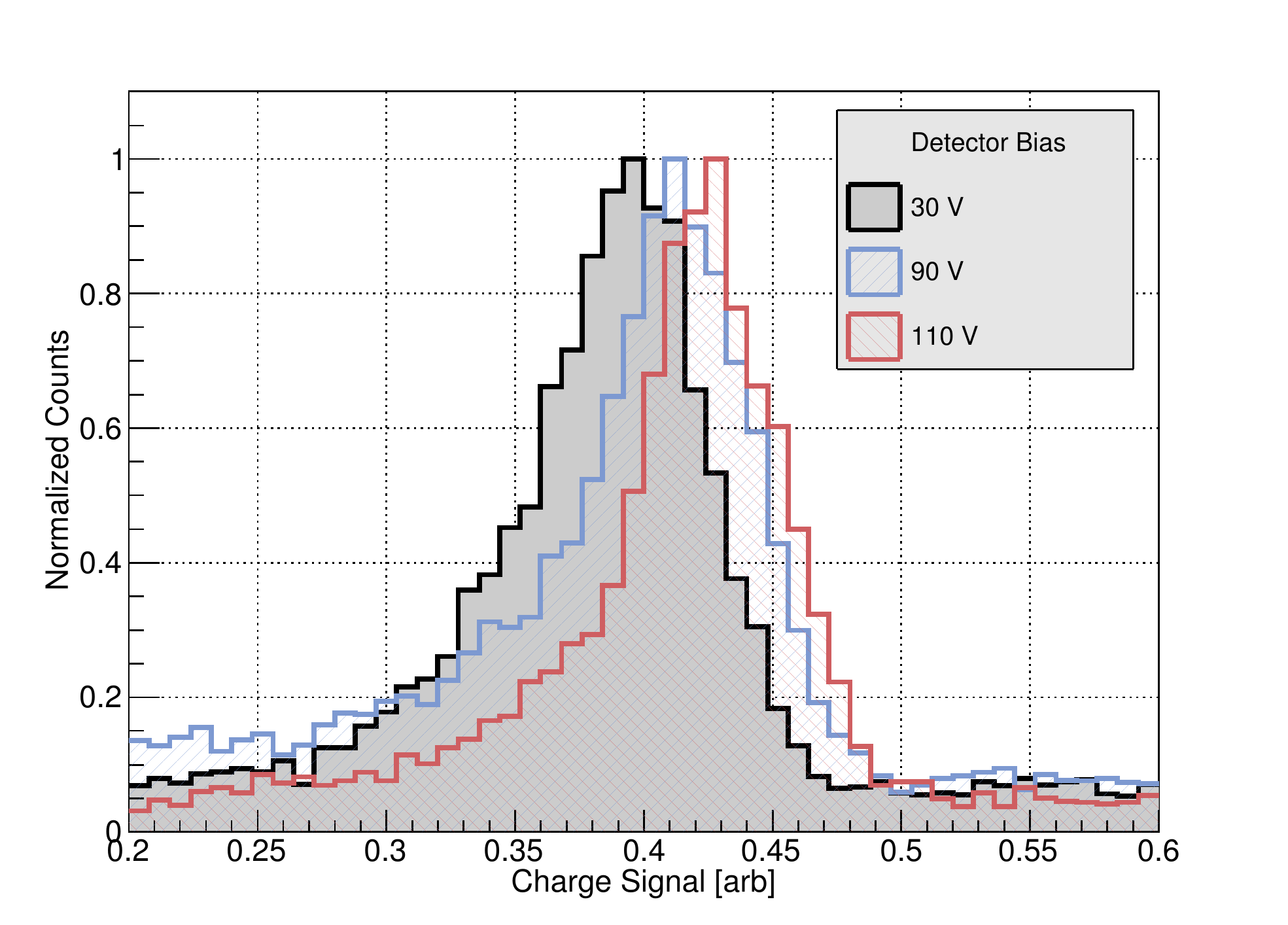}
    \caption{Shift of the 60 keV spectral line as measured by S1502 at various applied bias voltages. Fewer charges are collected at lower bias voltages because of trapping, recombination and diffusion effects, thus leading to lower charge collection efficiencies.}
    \label{fig:spectra_shift}
\end{figure}

\begin{figure}
	\centering
	\includegraphics[width=0.47\textwidth]{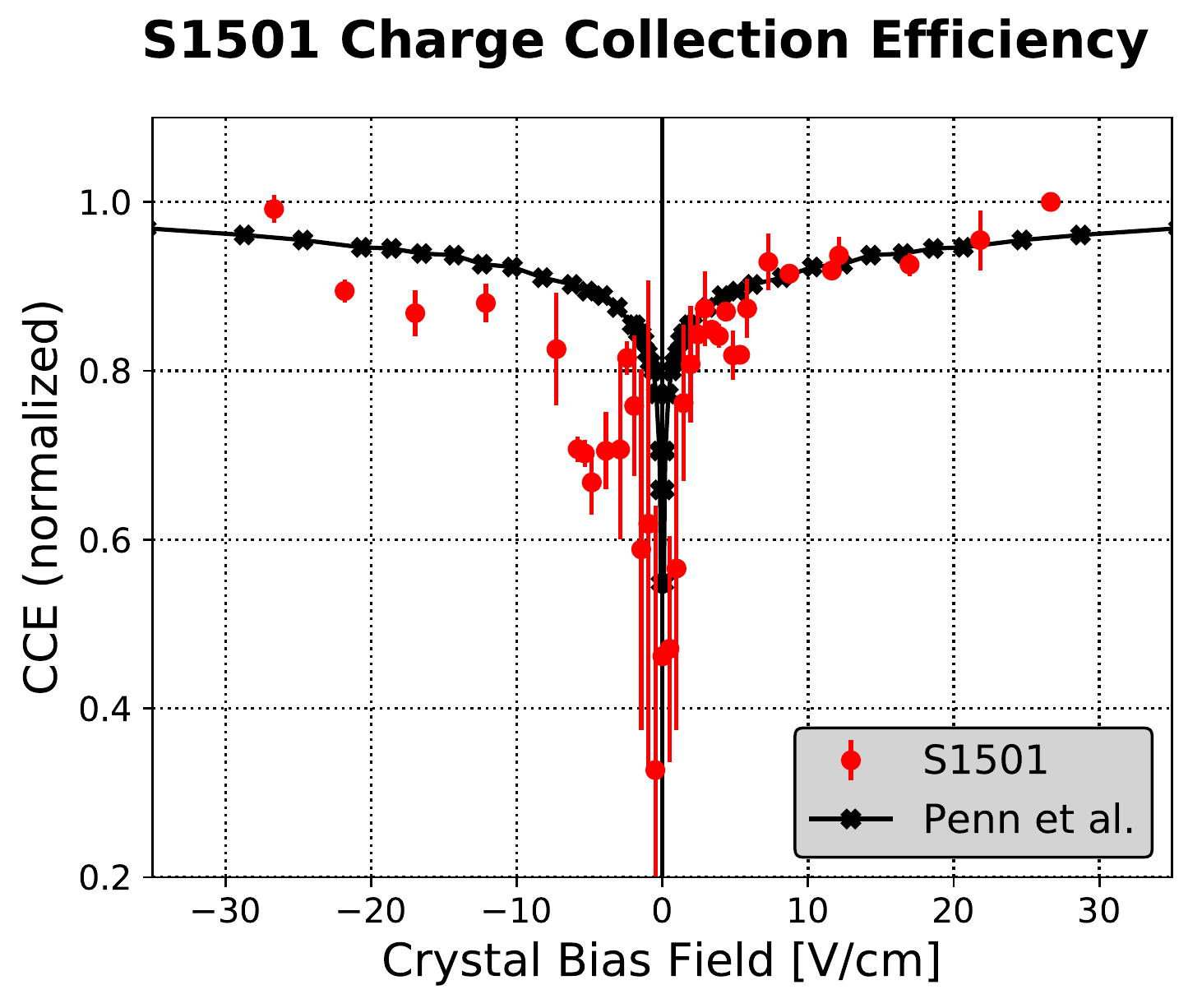}
	\includegraphics[width=0.49\textwidth]{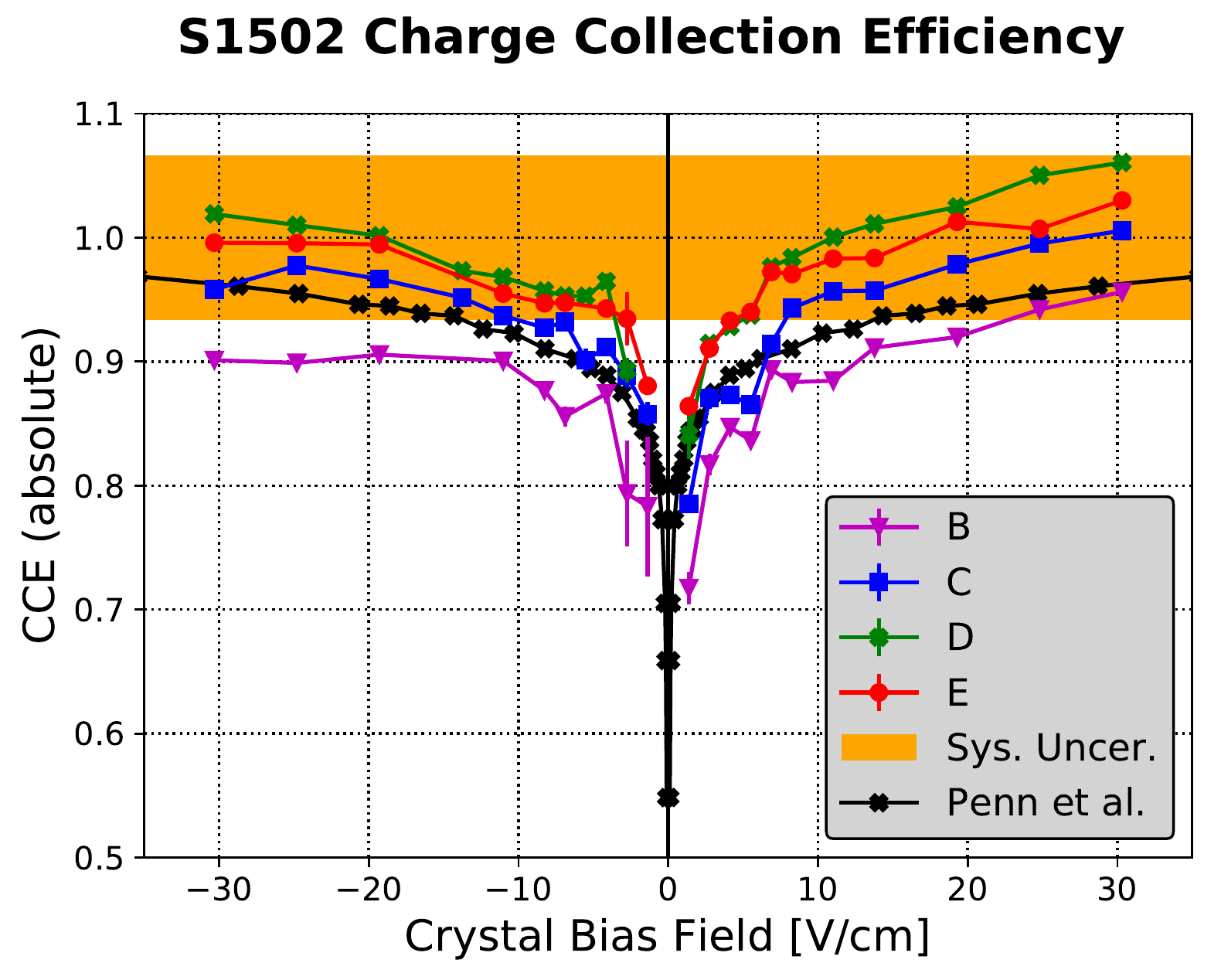}
	\caption{Charge collection efficiency as a function of applied bias. \textit{Left:} S1501 relative charge collection efficiency. Due to the lack of absolute calibration, this data has been normalized to have the highest collection efficiency = 1. \textit{Right:} S1502 charge collection efficiency curves showing lower efficiency at higher radii. The systematic uncertainty of the absolute charge collection efficiency scale is shown as an orange band. Previous results from a smaller (1 cm$\times$1 cm$\times$4.8 mm) Si device \cite{Penn1996} are also shown for comparison.}
	\label{fig:CCE}
\end{figure}

At low bias, some liberated carriers recombine or are trapped in impurities before fully traversing the crystal, thus reducing the charge signal. The charge collection efficiency (CCE) is defined as the ratio of measured induced charge on the readout electrode to that expected for a given event assuming no carriers are lost to these processes. As the bias voltage is increased, carriers are less likely to trap or recombine leading to increased CCE and thus larger signal amplitudes for a given interaction energy. This is demonstrated in Fig. \ref{fig:spectra_shift} where the measured signal from a calibration peak is observed to shift with changing bias voltage.

The average signal amplitude of the \textsuperscript{241}Am 60 keV calibration line was used to calculate the CCE as a function of bias field. For each channel with a calibration source, an energy spectrum was constructed from events which passed basic quality criteria such as good fits to the pulse template. In the multichannel detector, S1502, we further required that there be a pulse in only a single channel and that the other channel pulse amplitudes were consistent with noise. This ensured that we only considered events in which all carriers were drifted to a single channel. Additionally, the spectra consisted of only events from the first $\sim$2--10 minutes of data collected after applying the bias field to ensure that the applied field had not yet been reduced by time-dependent effects (discussed in Sec. \ref{sec:Neutralization}). The results of these measurements are shown in Fig. \ref{fig:CCE}. The left plot of Fig. \ref{fig:CCE} shows the CCE curve for the single channel detector, S1501. As discussed above, electronics issues prevented an absolute calibration of this device, so the curve is normalized to unity at the highest measured value. The right plot shows the CCE curves for the four inner channels of S1502. Here, the signal amplitudes have been calibrated absolutely. This calibration includes an overall uncertainty of the CCE scale primarily due to the error in the measured vacuum gap distance.

At high crystal bias fields, the full amount of liberated charge is measured and the CCE approaches 100\% within uncertainty. As the field is decreased, carriers are lost to trapping and recombination so the measured signal is observed to decrease, reducing CCE. With no bias field, the carriers have no net drift velocity and simply diffuse until they recombine or are trapped. In this case there is no measured signal. Previous measurements of Si CDMS detectors of both smaller thicknesses (5--33 mm) and radii (1--100 mm) \cite{Penn1996,Akerib:2005zy, bailey, Chagani:2014caa} have shown bias fields of $\sim$5 V/cm are required for full charge collection efficiency. Previous normalized CCE measurements of a square 1 cm $\times$ 1 cm, 4.8 mm thick Si detector \cite{Penn1996} are also shown in Fig. \ref{fig:CCE} for comparison. The results presented here for 150 mm diameter detectors demonstrate similar collection efficiency behavior and achieve 80--90\% CCE at crystal biases of 5--10 V/cm.

A simple model of carrier trapping to describe the shape of such CCE curves as a function of bias and crystal thickness was presented in Ref. \cite{Penn1996}. There it was found that, for high-resistivity Si devices less than 5 mm thick, charge loss was due to a combination of loss in the initial charge cloud and trapping while drifting the charge across the crystal. The model includes diffusion in the initial charge cloud with a length scale, $L$. All carriers that diffuse less than a field-dependent radius, parameterized as $bF^{-n}$, are trapped before the bias field, $F$, erodes the cloud. The population of carriers which survive the initial charge cloud recombination are then drifted across the crystal, attenuated by a field-dependent capture length, $cF^m$. The expression for CCE of a contact-free detector in terms of these parameters is then

\begin{multline}
CCE=\frac{2}{h+\kappa d}\left(1+\frac{b}{L}F^{-n}\right) exp\left[-\frac{b}{L}F^{-n}\right] \\
\times \left\{b F^{-n} +  cF^m \left(1-exp\left[-\frac{h-2 b F^{-n}}{2 c F^m}\right]\right)\right\}
\end{multline}

\begin{figure}
    \centering
    \includegraphics[width=0.8\textwidth]{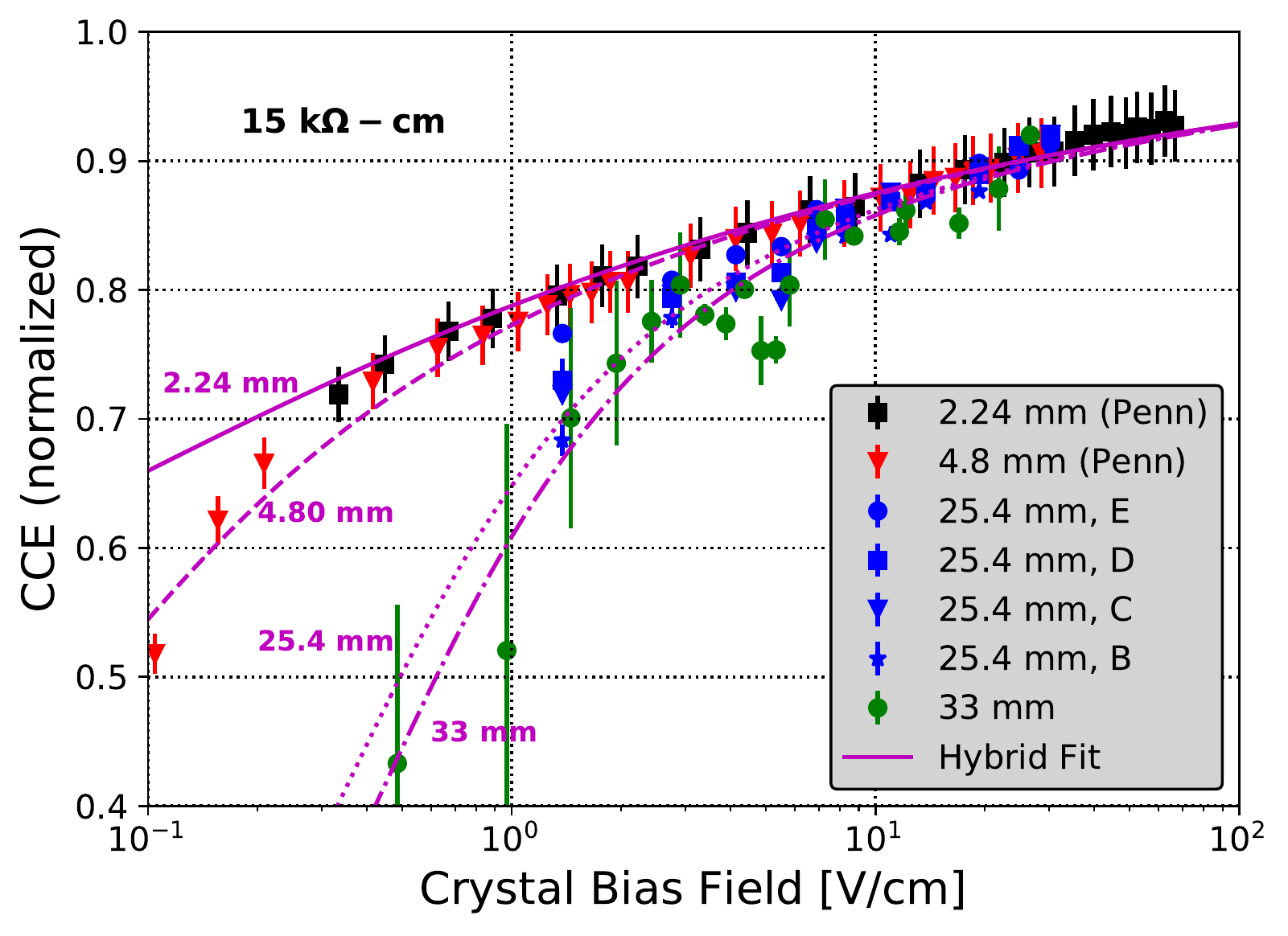}
    \caption{Fitting the trapping model of Ref. \cite{Penn1996} to combined CCE data.}
    \label{fig:CCE_fit}
\end{figure}

This same ``hybrid'' trapping model was used to simultaneously fit the data presented here and that for devices of similar resistivity presented in \cite{Penn1996}. Here, the measured data has been normalized to the extrapolated maximum CCE which introduces an additional $\sim$5$\%$ uncertainty. The resulting fit is shown in Fig. \ref{fig:CCE_fit} and the best-fit parameters can be found in Table \ref{tab:CCE_fit_params}. It is found that the length scale for trapping while drifting is $\sim$2 cm and increases approximately linearly with field. This indicates that, for devices less than a few cm thick, charge collection efficiency is dominated by trapping in the initial $\sim$0.7 mm charge cloud, before the carriers have drifted a significant distance from the initial event location.

\begin{table}[h]
\centering
\begin{tabular}{ |c|c|c| }
\hline
Parameter & Description & Best Fit \\
\hline
b & Max cloud radius at E=0 & 0.065 $\pm$ 0.016 cm \\
b/L & Ratio of b to diffusion length scale & 0.851 $\pm$ 0.017 \\
n & Max cloud radius inverse field exponent & 0.146 $\pm$ 0.008 \\
c & Average capture length at E=0 & 2.76 $\pm$ 0.171 cm \\
m & Capture length field exponent & 1.13 $\pm$ 0.081 \\
\hline
\end{tabular}
\caption{Best-fit parameters of hybrid trapping model to combined data set. Model and parameters are discussed in depth in Ref. \cite{Penn1996}.}
\label{tab:CCE_fit_params}
\end{table}

The S1502 measurements appear to show a modest decrease in CCE with radius. The inner channels, D and E, yield systematically higher CCE curves than the higher radius channels C and B. The outermost channel, A, did not have a calibration source and thus a CCE was not measured. It was, however, used for data selection criteria for channels as described above. This decrease of CCE with radius is due to either an increase in charge recombination/trapping or a decrease in induced ionization signal due to field non-linearities near the detector edges. Corrections of $\sim$5\% near the crystal edge are not unexpected given the detector geometry \cite{Nishiyama}. Work is currently being done to fully model the bias field, Ramo potentials, and charge transport in order to understand the radial difference in detector response in more detail.

\subsection{Crystal Neutralization}
\label{sec:Neutralization}

\begin{figure}
    \centering
    \includegraphics[width=0.9\textwidth]{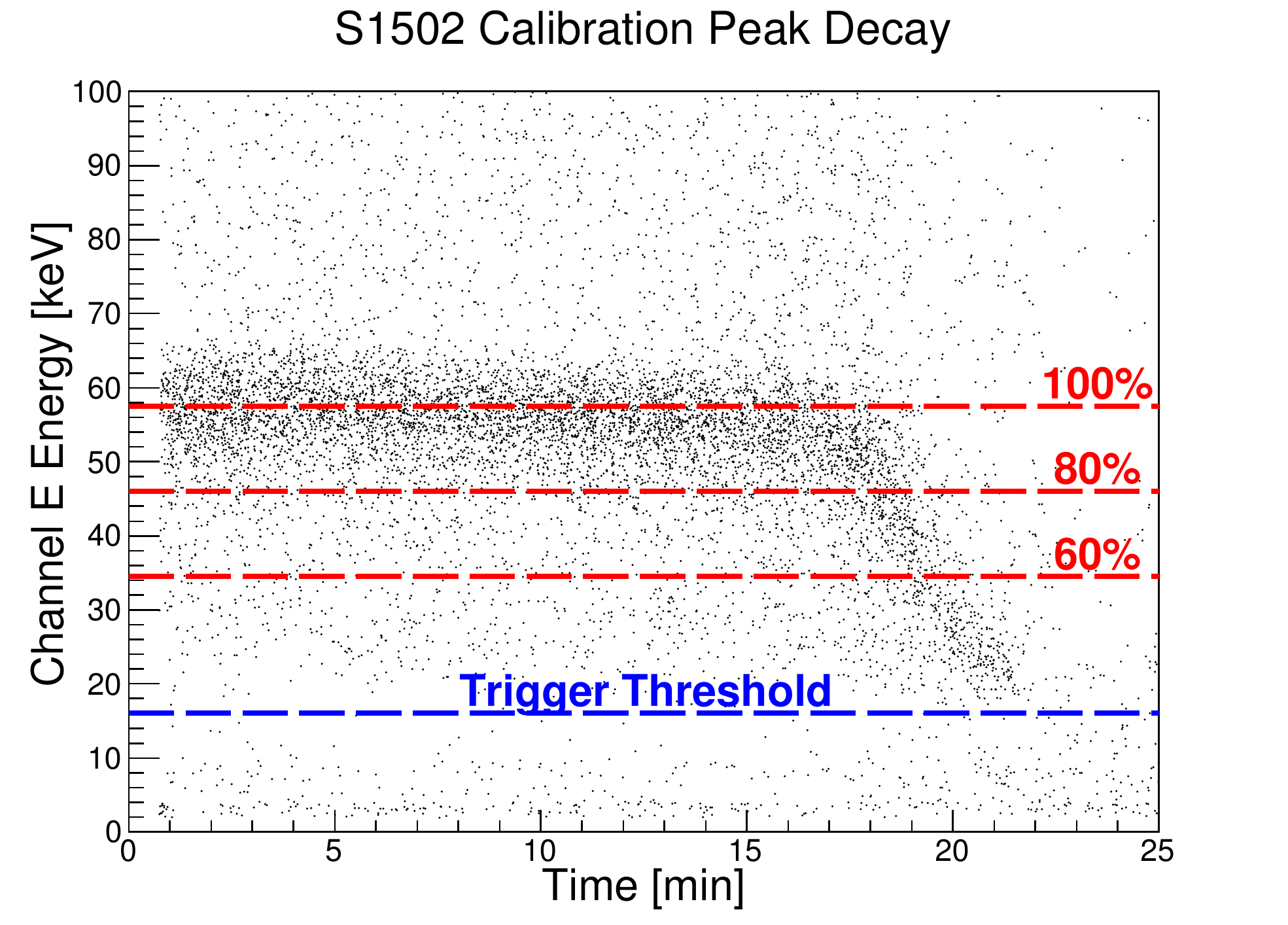}
    \caption{An example of the measured signal of a calibration peak in S1502 falling over time as liberated charge accumulates on the bare crystal face, thereby partly canceling the applied -25 V bias field. The times taken for a calibration peak to fall to 80\% and 60\% of its initial value are measured.}
    \label{fig:peak_v_time}
\end{figure}

As a detector is operated, charge trapping sites in the crystal gradually become filled and thus charged. These sites can interfere with the ionization signal, reducing its amplitude.  The contact-free design of these detectors added a further complication. Because one (both in the case of S1501) crystal face(s) was(were) electrically isolated from the bias electrodes, drifting charges were never actually able to be collected at an electrode, and they instead accumulated at the bare detector face(s). This accumulating charge caused a ``counter-bias'' field to develop over time during operation, reducing the effective bias field in the crystal bulk and lowering the charge collection efficiency. When either of these effects occurred, the crystal was said to have lost ``neutralization'' and needed to be reset.

The loss of neutralization was mitigated by periodically removing the bias field, grounding the electrodes, and illuminating the detector with pulses of LED light for several seconds in a process called ``flashing'' \cite{Chagani:2012zz}. The standard flashing pattern consists of powering the LED in 100 $\mu$s pulses at a rate of 200 Hz for 30 seconds. The photons produce free charge carriers in the crystal which then diffuse to overcharged regions neutralizing any bound space charge. If the crystal is already neutralized, the carriers are simply collected by the grounded electrode. In this manner, trapped charges are neutralized and the detector is reset. Flashing typically heats the fridge by $\sim$500 mK, so the detector is allowed to cool back to base temperature before the detector bias is applied and data collection is resumed.

\subsubsection{Accumulating Counter-Bias}
\label{sec:counter_bias}
The effect of the accumulating counter-bias is shown in the charge signal amplitude versus time plot in Fig. \ref{fig:peak_v_time} where the signal measured from a calibration peak is seen to decrease over time. As the counter-bias field grows, the effective bias field in the crystal falls and thus so does the charge collection efficiency, following the trends shown in Fig. \ref{fig:CCE}. When the effective bias field falls to $\sim$5 V/cm, the measured signals begin to rapidly decrease.

It has been observed that not only the calibration peak, but the entire measured energy spectrum falls at the same time and rate and that this occurs for all channels simultaneously in S1502. This implies that the dominant source of counter-bias charge is the spatially uniform background events, not the calibration sources (which are localized to one area of each readout channel).

From this model of neutralization loss the bias across the crystal should be (again ignoring fringing fields)
\begin{equation}
V_{cr} = \frac{h}{h+\kappa d}\left(V_{tot}-\frac{\sigma d}{\epsilon_0} \right)
\label{eq:counter_bias}
\end{equation}
where $\sigma$ is the accumulated surface charge density at the bare crystal face(s). This expression holds for both single and double vacuum interface devices (i.e. both S1501 and S1502) and assumes uniform $\sigma$. Here the accumulated surface charge provides the counter-bias, $V_{CB}$, which is defined as the effective reduction in the total applied detector bias.

\begin{equation}
V_{CB}=\frac{\sigma d}{\epsilon_0}
\end{equation}

A model has been developed to quantitatively understand this neutralization loss over time. We first assume there is a constant rate of energy deposition in the detector from ionizing radiation. Integrating the measured energy spectrum (adjusted for detector live time) yields a rate of 90.3 MeV/s deposited in the detector, mainly from background radiation. Assuming an average ionization energy of 3.8 eV / charge pair in Si \cite{PEHL196845}, we calculate the rate of carrier production in the crystal as $2.4 \times 10^7$ pairs/s. The measurements of CCE (Fig. \ref{fig:CCE}) then give the fraction of produced carriers which traverse the crystal bulk and accumulate on the bare crystal surface. These carriers contribute to the surface charge density, $\sigma$, increasing the counter-bias and thereby lowering CCE. In practice, the model iterates forward in small time steps calculating the charge accumulation rate, total amount of charge accumulated, and updated effective crystal field at each step.

\begin{figure}
    \centering
    \includegraphics[width=0.9\textwidth]{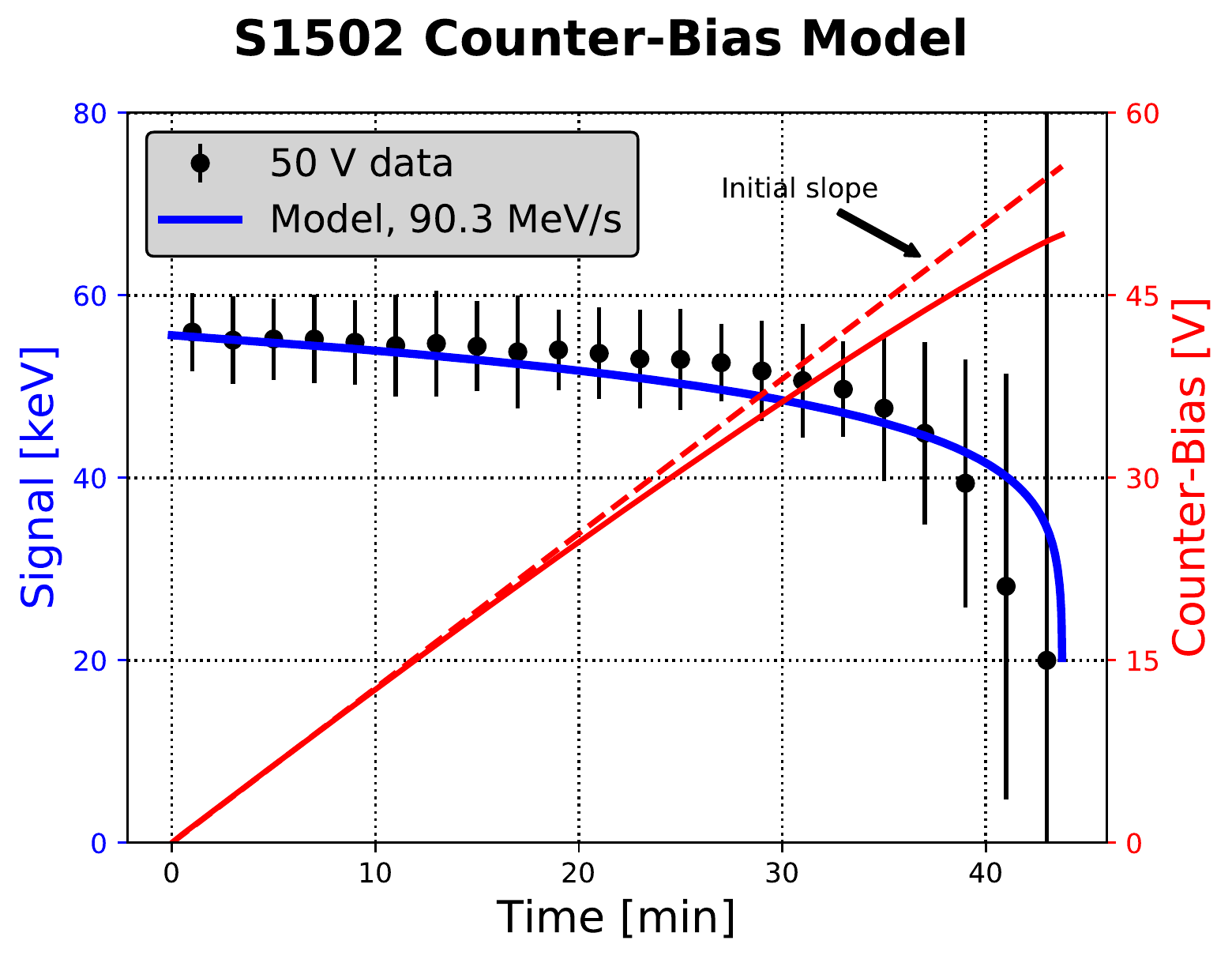}
    \caption{An example of the modeled counter-bias development and resulting calibration signal for the case of a total detector bias of 50 V and energy deposition rate in the detector of 90.3 MeV/s. The dashed line shows the initial counter-bias slope for illustration. The measured signal is due to 60 keV events with the effects of CCE applied. Shown for comparison are measured 60 keV peak signals.}
	\label{fig:r65model_eg}
\end{figure}

An example of this model's output is shown in Fig. \ref{fig:r65model_eg} where the counter-bias and 60 keV calibration line signal are plotted as a function of time. The counter-bias initially grows at a constant rate (dashed line) set by the initial CCE and the energy deposition rate. As the counter-bias becomes substantial, CCE falls, lowering both the observed signal and the rate at which carriers accumulate at the surface. Eventually the counter-bias is as large as the applied total bias, and the signal is lost. The model relies on an empirical parameterization of the average CCE curve which results in the slight overestimation of the signal for the example shown in Fig. \ref{fig:r65model_eg}.

\begin{figure}
    \centering
    \includegraphics[width=0.9\textwidth]{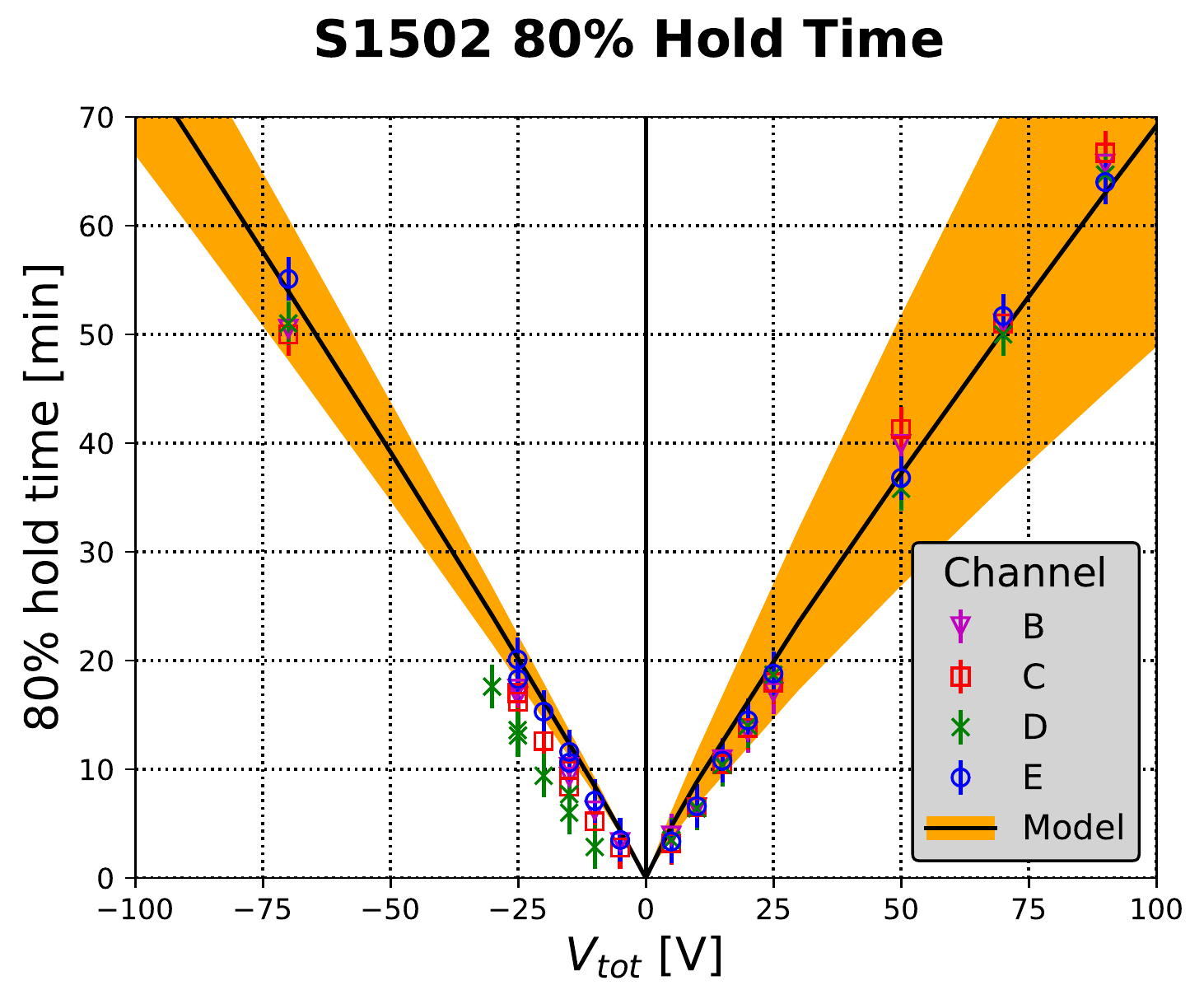}
    \caption{The measured 80\% signal hold time as a function of total applied detector bias. The results of a simple model of accumulating surface charge are seen to fit the data well. The model uncertainty is higher at positive biases due to larger variation in measured CCE curves at positive biases. The discrepancy in channel D at negative biases is believed to be due to a readout electronics issue.}
	\label{fig:r65ht80}
\end{figure}

To quantify this signal loss in data, we measured the time it took for the 60 keV line to fall to 80\% of its starting value (80\% hold time) in each channel of S1502 at several total detector bias voltages, $V_{tot}$. Fig. \ref{fig:r65ht80} shows this data as well as the 80\% hold times calculated by the model. This calculation used a parameterization of the average CCE vs. bias curve of the four channels. The orange band is the 1$\sigma$ uncertainty associated with the spread in individual channel CCE measurements. This simple model agrees well with most of the data, especially at high bias voltages. The only major discrepancy is for channel D at low, negative biases. The source of this deviation is not yet fully understood, but is likely related to the amplifier electronics on channel D, which were an earlier prototype of those used on the other channels. It should be noted that the model shown in Fig. \ref{fig:r65ht80} has no free parameters but agrees well with the measured data. This provides further evidence that the model of counter-bias due to accumulated surface charge is sufficient to describe the observed signal loss. 

We can also use this result to place an estimate on the amount of leakage current present in this device. A conservative estimate is to require the leakage current be lower than the average current due to the background event rate. This translates to a requirement that the current leakage through this detector be $\lesssim$3.8 pA. This limit is a factor of $\sim$10 larger than the leakage currently hampering some CDMS high-voltage detectors. Using such measurements of counter-bias over time may be a sensitive probe of low leakage current levels, provided the background event rate is low. This technique is powerful because it effectively integrates the current over long time scales which could allow the measurement of low, constant leakage currents.

\subsubsection{Detector Reset with LEDs}
After the detector signal has been degraded by the counter-bias field, it needs to be reset by eliminating overcharged areas and ``neutralizing'' the crystal. We tested neutralization in S1502 using varied amounts of flashing from the different LEDs (IR mounted on bias electrode or on detector housing, UV mounted on bias electrode) while the detector was grounded. The IR LEDs had a spectrum peaked at 940 nm (1.31 eV) which is above the Si indirect band gap (1.12 eV) but below the direct band gap (3.4 eV). The UV LED spectrum was peaked at 310 nm (4.0 eV) which is higher than the direct band gap. After each flashing, a 10 V total detector bias was applied for data collection. With a standard flash, this 10 V bias provided an 80\% hold time of 6 minutes. We considered this condition as ``full neutralization''. We varied the following: which LEDs were used, total flash time, LED pulse rate, LED pulse width.

\begin{figure}
    \centering
    \includegraphics[width=0.7\textwidth]{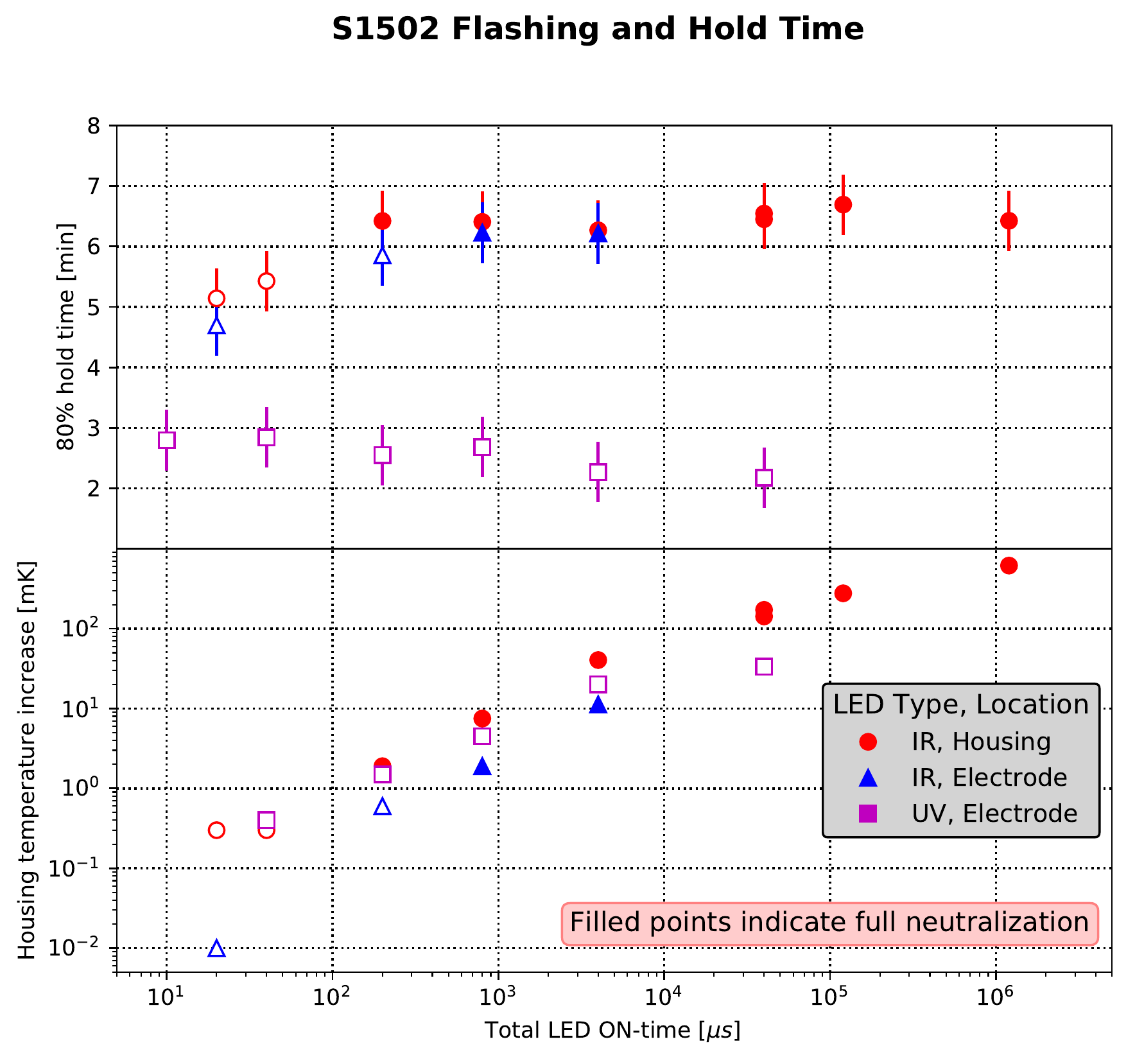}
	\caption{\textit{Upper:} IR LEDs on both the electrode and the housing provided full neutralization over a wide range of total on-times. Housing LEDs performed a little better, providing full neutralization at shorter ON-times. UV LEDs did not neutralize any better than no flashing. \textit{Lower:} Temperature increase of the housing was a good predictor of whether the detector neutralized completely for IR LEDs. Although they seemed to be providing similar power, the UV LEDs were not observed to help neutralize the detector.}
    \label{fig:r65_flashing_compare}
\end{figure}

Fig. \ref{fig:r65_flashing_compare} summarizes the results of these flashing tests as a function of total time the LEDs were on during flashing. For each type of LED, this is related to the total ionizing energy delivered to the detector. The detector housing temperature increase is used to compare the different LED power outputs. The lower plot of Fig. \ref{fig:r65_flashing_compare} demonstrates that the total ``LED ON-time'' was well correlated with detector housing temperature increase and thus emitted power. The detector is well coupled thermally to this housing, so this is a good proxy for relative power delivered to the detector.

The upper plot indicates that the IR LEDs on both the bias electrode and the detector housing provided full neutralization over a wide range of total ON-times. The IR LEDs on the detector housing performed a little better, providing full neutralization at lower ON-times. For the range of flashing tested, the UV LEDs, located on the bias electrode, did not neutralize any better than simply grounding the detector without flashing. The lower plot indicates that for the IR LEDs the temperature increase of the tower was a good predictor of whether the detector neutralized completely. Although the UV LEDs seemed to be emitting the same amount of power, they did not help neutralize the detector.

Another important observation was that grounding with no flashing at all also neutralized the detector. Grounding for 5 minutes (which is close to the standard fridge cool-down time after flashing) produced partial neutralization (4 minute 80\% hold time); grounding for 25 minutes produced full neutralization. Even without the LEDs, there is a constant background event rate in the detector depositing $\sim$90 MeV/s which liberates carriers that also help neutralize the detector.

The difference in neutralization properties of the IR and UV LEDs may be related to the penetration depths of the different photons into the Si detector. If the photons are absorbed very near the crystal surface, the carriers produced may be less likely to neutralize the crystal by diffusing into the bulk material. At room temperature, the absorption coefficients for UV (310 nm) and IR (940 nm) light are ${\sim}10^6$ cm\textsuperscript{-1} and ${\sim}10^3$ cm\textsuperscript{-1} respectively \cite{sze}. The UV energy is greater than the Si direct bandgap so it is absorbed much more readily, with most of the UV energy being deposited within $\sim$10 nm of the surface. The IR light, on the other hand, can only liberate carriers across the indirect bandgap of Si which also requires a lattice phonon, making it a less frequent interaction and increasing the IR penetration depth. At cryogenic temperatures, the decreased phonon population further lowers this likelihood, increasing the IR penetration depth to more than 10 \mu m. Perhaps this difference in photon absorption profile means carriers freed by UV must diffuse further to neutralize the crystal bulk, making the UV process less efficient.

The flashing and cool down process for these devices generally took 30 minutes. Given hold times of 10 min -- 1hr, this was a significant component of the detector's operational duty cycle at the test facility. The hold time is driven by charge accumulation on the contact-free crystal surface(s) due to events in the detector. If operated in a low-background environment, such as that for a dark matter search where the energy deposition rate is several orders of magnitude lower, the detector hold time would be proportionally longer. Such a detector could easily achieve duty cycles of \textgreater 90\% if operated with $V_{tot} \gtrsim$ 50 V in a low-background facility such as SNOLAB.
\section{Conclusions}
The prototype contact-free detector designs with 150 mm diameter Si crystals show sufficient charge collection efficiency and crystal neutralization for operation as future generation cryogenic particle detectors. Contact-free biasing and readout has also been successfully demonstrated. High event rates can cause counter-bias accumulation which reduces the measured signal, but this can be mitigated by periodic LED flashing. This effect would also be greatly reduced when operating underground in low background environments. It was found that IR photons neutralized the detectors more efficiently than UV photons, perhaps due to the greater penetration of IR in cryogenic Si. If the background event rate is high enough, simply grounding the detector for a period of time can be just as effective at resetting the device.
These promising results call for the design, fabrication, and implementation of phonon sensors with these large crystals as the next step in developing this detector technology.

\section*{Acknowledgments}
We would like to thank P. Nelsen in the UMN CSE Shop who machined several important detector components. We would also like to thank A. Villano for useful discussions and guidance. This work was supported by the DOE grants DE-SC0012294 and DE-SC0011824.


\section*{References}

\bibliographystyle{apsrev-custom}

\bibliography{bibliography}

\end{document}